\documentclass[epj]{svjour}
\usepackage{graphics}
\begin{document}
\title{Prolate dominance and prolate-oblate shape transition in the proxy-SU(3) model}

\author{Dennis Bonatsos\inst{1} \and I. E. Assimakis \inst{1} \and N. Minkov\inst{2} \and Andriana Martinou\inst{1} \and \underline{S. Sarantopoulou}\inst{1} \and R. B. Cakirli\inst{3} \and R. F. Casten\inst{4,5} \and
K. Blaum\inst{6}} % etc
%}                     % Do not remove
\institute{Institute of Nuclear and Particle Physics, National Centre for Scientific Research 
``Demokritos'', GR-15310 Aghia Paraskevi, Attiki, Greece
\and Institute of Nuclear Research and Nuclear Energy, Bulgarian Academy of Sciences, 72 Tzarigrad Road, 1784 Sofia, Bulgaria
\and Department of Physics, University of Istanbul, Istanbul, Turkey
\and Wright Laboratory, Yale University, New Haven, Connecticut 06520, USA
\and Facility for Rare Isotope Beams, 640 South Shaw Lane, Michigan State University, East Lansing, MI 48824 USA
\and  Max-Planck-Institut f\"{u}r Kernphysik, Saupfercheckweg 1, D-69117 Heidelberg, Germany}

\date{Received: date / Revised version: date}
\abstract{
Using a new approximate analytic parameter-free proxy-SU(3) scheme, we make simple predictions  for
the global feature of prolate dominance in deformed nuclei and the locus of the prolate-oblate shape transition and compare these with empirical data.
}
 % end of PACS codes
 %end of abstract
%
\maketitle
\section{Introduction}
\label{intro}

The dominance of prolate (rugby ball) over oblate (pancake) shapes in the ground states of even-even nuclei is still a puzzling open problem in nuclear structure \cite{HM}. Furthermore, there is experimental evidence for a prolate-to-oblate shape/phase transition in the rare earth region 
around neutron number $N=116$ \cite{Alkhomashi,Namenson,Wheldon,Podolyak,Linnemann}, in agreement 
with predictions by microscopic calculations \cite{Sarriguren,Robledo,Nomura83,Nomura84}. We are going to consider these two issues in the framework 
of the proxy-SU(3) model, a new approximate symmetry scheme applicable in medium-mass and heavy deformed nuclei \cite{PRC1,PRC2}. The basic features and the theoretical foundations of proxy-SU(3),
as well as some first applications for making predictions for ground state properties of even-even deformed nuclei, have been described in Refs. \cite{IoBonat,IoAssim,IoMartinou}, to which the reader is referred. 

\section{Particle-hole symmetry breaking}

We are going to show that the prolate vs. oblate dominance in the deformed shapes of even-even nuclei 
is rooted in the breaking of the particle-hole symmetry in the shell picture of the nucleus.
A glance at the level schemes of the Nilsson model \cite{Nilsson1,Nilsson2} suffices in order to see that the first 
few orbitals lying in the beginning of a shell carry values of quantum numbers completely different 
from the last few orbitals at the end of the shell. In the neutron 50-82 major shell, for example, 
and at a moderate deformation $\epsilon=0.3$, the three orbitals at the beginning of the shell 
are 1/2[431], 1/2[420], and 1/2[550], while the last three orbitals below the top of the shell
are 11/2[505], 3/2[402], and 1/2[400], with the usual Nilsson notation $K[N n_z \Lambda]$ being used, 
where $N$ is  the number of oscillator quanta, $n_z$ is the number of oscillator quanta  along the cylindrical symmetry axis $z$, $\Lambda$ is the $z$-projection  of the orbital angular momentum, 
and $K$ is the $z$-projection of the total angular momentum.

%%%%%%%%%%%%%%%%%%%%%%%%%%% FIG. 1 %%%%%%%%%%%%%%%%%%%%%%%%%%%%%%%%%%%%%
\begin{figure}[htb]

\resizebox{0.49\textwidth}{!}{%
\includegraphics{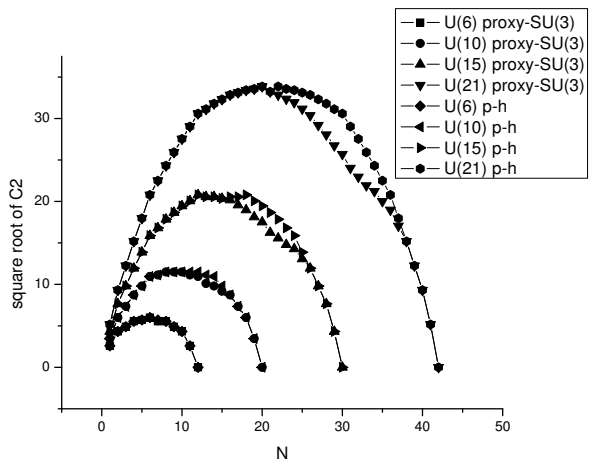}
}

\caption{Values of the square root of the second order Casimir operator of SU(3), obtained from 
Eq.~(1), vs. particle number N, for different shells, obtained through proxy-SU(3) or through the particle-hole symmetry assumption.}

\end{figure}

In the framework of the proxy-SU(3) the particle-hole symmetry breaking can be seen in Table 1,
in which the SU(3) irreducible representations (irreps) \cite{Elliott1,Elliott2} corresponding to the highest weight state for a given number of particles (protons or neutrons) 
are shown. The harmonic oscillator shell, possessing a U(N) symmetry, corresponding to each nuclear shell within the proxy-SU(3) scheme is also shown in Table 1. The highest weight states have been obtained by using the code UNTOU3 \cite{code}. We see that the particle-hole symmetry is valid 
only up to 4 particles and 4 holes, while it is broken in the middle of each shell. In the $sd$ 
shell however, the particle-hole symmetry breaking is absent for even particle numbers and it appears 
only for one odd particle number, due to the small size of the shell. It should be noticed that 
no particle-hole symmetry breaking occurs if, instead of the highest weight irreps, one considers the irreps possessing the highest eigenvalue of the second order Casimir operator of SU(3) \cite{IA} 
 \begin{equation}\label{C2} 
 C_2(\lambda,\mu)= {2 \over 3} (\lambda^2+\lambda \mu + \mu^2+ 3\lambda +3 \mu). 
\end{equation}
The difference can be seen in Fig. 1, where the values of the square root of the Casimir operator are shown for the two cases in discussion.  
The particle-hole symmetry breaking can also be seen in Table 5 of Ref. \cite{pseudo1}, where the odd numbers of particles in the U(10) shell are reported. 

%%%%%%%%%%%%%%%%%%%%%%%% Table 1  %%%%%%%%%%%%%%%%%%%%%%%%%%%%%%%
\begin{table*}[htb]

\caption{Comparison between SU(3) irreps for U(6), U(10), U(15), and U(21), obtained by the code UNTOU3 \cite{code},  contained in the relevant U(n) irrep  for M valence protons or M valence neutrons.
Above the U(n) algebra, the relevant shell of the shell model  and the corresponding proxy-SU(3) shell are given.
The highest weight SU(3) irreps, given in the columns labelled by hw, are compared to the  SU(3) irreps with the highest eigenvalue of the second order Casimir operator of SU(3),
given in the columns labelled by C. Irreps breaking the particle-hole symmetry in the hw columns are indicated by boldface characters. Taken from Ref. \cite{PRC2}.
}

\bigskip
\begin{tabular}{ r l r r r r r r r r  }

\hline

\hline
   &             & 8-20 & 8-20 & 28-50 & 28-50& 50-82 & 50-82 & 82-126& 82-126\\
   &             & sd   &  sd  & pf    &  pf  & sdg   &  sdg  &  pfh  & pfh   \\
M  & irrep       & U(6) & U(6) & U(10) & U(10)& U(15) & U(15) & U(21) & U(21) \\
   &             & hw   & C    & hw    &  C   & hw    &  C    & hw    &   C   \\
 0 &             &(0,0) &(0,0) &(0,0)  &(0,0) &(0,0)  &(0,0)  &(0,0)  &(0,0)  \\  
 1 & [1]         &(2,0) &(2,0) & (3,0) &(3,0) & (4,0) &(4,0)  & (5,0) &(5,0)  \\
 2 & [2]         &(4,0) &(4,0) & (6,0) &(6,0) & (8,0) &(8,0)  &(10,0) &(10,0) \\
 3 & [21]        &(4,1) &(4,1) & (7,1) &(7,1) &(10,1) &(10,1) &(13,1) &(13,1) \\
 4 & [$2^2$]     &(4,2) &(4,2) & (8,2) &(8,2) &(12,2) &(12,2) &(16,2) &(16,2) \\
 5 & [$2^2$1]    &(5,1) &(5,1) &(10,1) &(10,1)&(15,1) &(15,1) &(20,1) &(20,1) \\
 6 & [$2^3$]     &(6,0) & (0,6)&(12,0) &(12,0)&(18,0) &(18,0) &(24,0) &(24,0) \\
 7 & [$2^3$1]  &{\bf(4,2)}&(1,5)&(11,2)&(11,2)&(18,2) &(18,2) &(25,2) &(25,2) \\
 8 & [$2^4$]     &(2,4) & (2,4)&(10,4) &(10,4)&(18,4) &(18,4) &(26,4) &(26,4) \\
 9 & [$2^4$1]    &(1,4) & (1,4)&(10,4) &(10,4)&(19,4) &(19,4) &(28,4) &(28,4) \\
10 & [$2^5$]     &(0,4) & (0,4)&(10,4) &(4,10)&(20,4) &(20,4) &(30,4) &(30,4) \\
11 & [$2^5$1]   &(0,2)&(0,2)&{\bf(11,2)}&(4,10)&(22,2)&(22,2) &(33,2) &(33,2) \\
12 & [$2^6$]    &(0,0)&(0,0)&{\bf(12,0)}&(4,10)&(24,0)&(24,0) &(36,0) &(36,0) \\
13 & [$2^6$1]   &     &     &{\bf(9,3)} &(2,11)&(22,3)&(22,3) &(35,3) &(35,3) \\
14 & [$2^7$]    &     &     &{\bf(6,6)} &(0,12)&(20,6)&(20,6) &(34,6) &(34,6) \\
15 & [$2^7$1]   &     &     &{\bf(4,7)} &(1,10)&(19,7)& (7,19)&(34,7) &(34,7) \\
16 & [$2^8$]    &  &      & (2,8) & (2,8)&{\bf(18,8)} & (6,20)&(34,8) &(34,8) \\
17 & [$2^8$1]   &  &      & (1,7) & (1,7)&{\bf(18,7)} & (3,22)&(35,7) &(35,7) \\
18 & [$2^9$]    &  &      & (0,6) & (0,6)&{\bf(18,6)} & (0,24)&(36,6) &(36,6) \\
19 & [$2^9$1]   &  &      & (0,3) & (0,3)&{\bf(19,3)} & (2,22)&(38,3) &(38,3) \\
20 & [$2^{10}$] &  &      & (0,0) & (0,0)&{\bf(20,0)} & (4,20)&(40,0) &(40,0) \\
21 & [$2^{10}$1]&  &      &       &      &{\bf(16,4)} & (4,19)&(37,4) & (4,37) \\
22 & [$2^{11}$] &  &      &       & &{\bf(12,8)} & (4,18)&{\bf(34,8)} &(0,40)\\
23 & [$2^{11}$1]&  &      &       & &{\bf(9,10)} & (2,18)&{\bf(32,10)}&(3,38)\\
24 & [$2^{12}$] &  &      &       & &{\bf(6,12)} & (0,18)&{\bf(30,12)}&(6,36)\\
25 & [$2^{12}$1]&  &      &       & &{\bf(4,12)} & (1,15)&{\bf(29,12)}&(7,35)\\
26 & [$2^{13}$] &      &      &   &      &(2,12) & (2,12)&{\bf(28,12)}&(8,34)\\
27 & [$2^{13}$1]&      &      &   &      &(1,10) & (1,10)&{\bf(28,10)}&(7,34)\\
28 & [$2^{14}$] &      &      &   &      & (0,8) & (0,8) &{\bf(28,8)} &(6,34)\\
29 & [$2^{14}$1]&      &      &   &      & (0,4) & (0,4) &{\bf(29,4)} &(3,35)\\
30 & [$2^{15}$] &      &      &   &      & (0,0) & (0,0) &{\bf(30,0)} &(0,36)\\
31 & [$2^{15}$1]&      &      &   &      &       &       &{\bf(25,5)} &(2,33)\\
32 & [$2^{16}$] &      &      &   &      &       &       &{\bf(20,10)}&(4,30)\\
33 & [$2^{16}$1]&      &      &   &      &       &       &{\bf(16,13)}&(4,28)\\
34 & [$2^{17}$] &      &      &   &      &       &       &{\bf(12,16)}&(4,26)\\
35 & [$2^{17}$1]&      &      &   &      &       &       &{\bf(9,17)} &(2,25)\\ 
36 & [$2^{18}$] &      &      &   &      &       &       &{\bf(6,18)} &(0,24)\\
37 & [$2^{18}$1]&      &      &   &      &       &       &{\bf(4,17)} &(1,20)\\
38 & [$2^{19}$] &      &      &        &      &       &       &(2,16) &(2,16)\\
39 & [$2^{19}$1]&      &      &        &      &       &       &(1,13) &(1,13)\\
40 & [$2^{20}$] &      &      &        &      &       &       &(0,10) &(0,10)\\
41 & [$2^{20}$1]&      &      &        &      &       &       &(0,5)  &(0,5) \\
42 & [$2^{21}$] &      &      &        &      &       &       &(0,0)  &(0,0) \\
\hline

\end{tabular}

\end{table*}

\section{Prolate over oblate dominance}

Using the group theoretical results reported in Table 1 for the rare earths with 50-82 protons 
and 82-126 neutrons, one obtains for the highest weight irrep corresponding to each nucleus 
the irreps shown in Table 2. Similar results for rare earths with 50-82 protons and 50-82 neutrons 
are shown in Table 3.  

As it was mentioned in Ref.  \cite{IoMartinou}, 
a connection between the collective variables $\beta$ and $\gamma$ of the collective model \cite{BM}
and the quantum numbers $\lambda$ and $\mu$ characterizing the irreducible represention 
$(\lambda,\mu)$ of SU(3) \cite{Elliott1,Elliott2} is known \cite{Castanos,Park}, based on the fact that the invariant quantities of the two theories should posses the same values,
the relevant equation for $\gamma$ being \cite{Castanos,Park}
\begin{equation}\label{g1}
\gamma = \arctan \left( {\sqrt{3} (\mu+1) \over 2\lambda+\mu+3}  \right). 
\end{equation}
From this equation one easily sees that irreps with $\lambda > \mu$ correspond to prolate shapes
with $0 < \gamma < 30^{\rm{o}}$, while irreps with $\lambda < \mu$ correspond to oblate shapes
with $30^{\rm{o}} < \gamma < 60^{\rm{o}}$. 

As a result, in Table~2 we see that most nuclei possess prolate ground states, with the exception 
of a few nuclei appearing simultaneously just below the $Z=82$ shell closure and just below  the
$N=126$ shell closure. Similar conclusions are drawn from Table~3, where a few oblate nuclei 
are predicted just below the $Z=82$ shell closure and just below  the
$N=82$ shell closure. 

\section{The prolate-to-oblate shape/phase transition}

In Table~2 it appears that a prolate-to-oblate shape/phase transition occurs at $N=116$, 
while in Table~3  a similar transition is predicted at $N=72$. The latter transition 
lies far away from the experimentally accessible region, but the first one is supported by existing
experimental data. In particular

1) In the W series of isotopes, $^{190}$W, which has $N=116$,  has been suggested as the lightest oblate isotope \cite{Alkhomashi}.   

2) In the Os series of isotopes, $^{192}$Os, which has $N=116$, has been suggested as lying at the 
shape/phase transition point \cite{Namenson,Wheldon}, with $^{194}$Os \cite{Namenson,Wheldon} 
and $^{198}$Os \cite{Podolyak} found to possess an oblate character. 

3) In the chain of even nuclei (differing by two protons or two neutrons) $^{180}$Hf, $^{182-186}$W, $^{188-192}$Os, $^{194,196}$Pt, $^{198,200}$Hg,
considered in Ref. \cite{Linnemann}, the collection of experimental data used 
suggests the the transition from prolate to oblate 
shapes occurs between $^{192}$Os and $^{194}$Pt, both of them possessing $N=116$.  

Furthermore, the present findings are in agreement with recent 
self-consistent Skyrme Hartree-Fock plus BCS calculations \cite{Sarriguren} and Hartree-Fock-Bogoliubov calculations \cite{Robledo,Nomura83,Nomura84} studying the structural evolution in neutron-rich Yb, Hf, W, Os, and Pt isotopes, reaching the conclusion that $N=116$ nuclei in this region can be identified as the transition points between prolate and 
oblate shapes. 

\section{Conclusions}

We have shown that the prolate over oblate dominance in the ground states of deformed 
even-even nuclei is predicted by the proxy-SU(3) scheme, based on the breaking of the 
particle-hole symmetry within each nuclear shell. Furthermore, we have seen that the proxy-SU(3) symmetry suggests $N=116$ as the point of the prolate-to-oblate shape/phase transition,
in agreement with existing exprerimental evidence \cite{Alkhomashi,Namenson,Wheldon,Podolyak,Linnemann} and microscopic calculations
\cite{Sarriguren,Robledo,Nomura83,Nomura84}.

%%%%%%%%%%%%%% TABLE 2trunc %%%%%%%%%%%%%%%%%%%%%%%%%%%%%%%%%%%%%%%%%%%%%%%%%%%%%

\begin{table*}[htb]

\caption{Most leading SU(3) irreps \cite{Elliott1,Elliott2} for nuclei with protons in the 50-82 shell 
and neutrons in the 82-126 shell.  Nuclei are divided into five groups: 1) Nuclei with $R_{4/2}=E(4^+_1)/E(2_1^+) \geq 2.8$ are indicated by boldface numbers. 
2) Nuclei with $2.8 > R_{4/2} \geq 2.5$ are denoted by *. 3) A few nuclei with  $R_{4/2}$ ratios slightly below 2.5,  shown for comparison, are labelled by **. 
4)  For any other nuclei with $R_{4/2}<2.5$, no irreps are shown. 5) Nuclei for which the $R_{4/2}$ ratios are still unknown \cite{ENSDF} are shown 
using normal fonts and without any special signs attached . Underlined  irreps correspond to oblate shapes.
Based on Ref. \cite{PRC2}.
}

\bigskip

\begin{tabular}{ r  |  r r r r r r r r r r r }

 &  Ce    &  Nd    &  Sm    &   Gd   &  Dy    &  Er    &  Yb    &  Hf    &   W    &   Os   &  Pt     \\
   $Z$   & 58     & 60     & 62     & 64     & 66     & 68     & 70     & 72     & 74     & 76     & 78      \\

\hline
122 &(18,14) &(20,14) &(24,14) &(20,16) &(18,18) &(18,16) &(20,10) &(12,18) &(6,22)  &(2,22)  &(0,18)    \\
120 &(20,20) &(22,20) &(26,16) &(22,22) &$\underline{(20,24)}$ &$\underline{(20,22)}$ &(22,16) &$\underline{(14,24)}$ &$\underline{(8,28)}$  
&$\underline{(4,28)*}$ &$\underline{(2,24)**}$ \\
118 &(24,22) &(26,22) &(30,18) &(26,24) &(24,16) &(24,24) &(26,18) &$\underline{(18,26)}$ &$\underline{(12,30)}$ &$\underline{(8,30)*}$ &$\underline{(6,26)**}$ \\
116  &(30,10) &(32,10) &(36,6)  &(32,12) &(30,14) &(30,12) &(32,6)  &(24,14) &$\underline{(18,28)*}$&  $\underline{\bf{(14,28)}}$&$\underline{(12,24)**}$\\
114 &(38,14) &(40,14) &(44,10) &(40,16) &(38,18) &(38,16) &(40,10) &(32,18) &\bf{(26,22)}&\bf{(22,22)}&(20,18)**\\
112 &(48,4)  &(50,4)  &(54,0)  &(50,6)  &(48,8)  &(48,6)  &(50,0)  &\bf{(42,8)}&\bf{(36,12)}&\bf{(32,12)}&(30,8)** \\
110 &(46,12) &(48,12) &(52,8)  &(48,14) &(46,16) &(46,14) &(48,8)  &\bf{(40,16)}&\bf{(34,20)}&\bf{(30,20)}&(28,16)* \\
108 &(46,16) &(48,16) &(52,12) &(48,18) &(46,20) &(46,18) &\bf{(48,12)}&\bf{(40,20)}&\bf{(34,24)}&\bf{(30,24)}&(28,20)* \\
106 &(48,16) &(50,16) &(54,12) &(50,18) &(48,20) &\bf{(48,18)}&\bf{(50,12)}&\bf{(42,20)}&\bf{(36,24)}&\bf{(32,24)}&(30,20)* \\
104 &(52,12) &(54,12) &(58,8)  &(54,14) &\bf{(52,16)}&\bf{(52,14)}&\bf{(54,8)}&\bf{(46,16)}&\bf{(40,20)}&\bf{(36,20)} &(34,16)* \\
102 &(58,4)  &(60,4)  &(64,0)  &\bf{(60,6)}&\bf{(58,8)}&\bf{(58,6)}&\bf{(60,0)}&\bf{(52,8)}&\bf{(46,12)}&\bf{(42,12)}&(40,8)*  \\
100 &(54,10) &(56,10) &(60,6)  &\bf{(56,12)}&\bf{(54,14)}&\bf{(54,12)}&\bf{(56,6)}&\bf{(48,14)}&\bf{(42,18)}&\bf{(38,18)}&(36,14)* \\
98 &(52,12) &(54,12) &\bf{(58,8)}&\bf{(54,14)}&\bf{(52,16)}&\bf{(52,14)}&\bf{(54,8)}&\bf{(46,16)}&\bf{(40,20)}&(36,20)*&   \\
96 &(52,10) &{\bf (54,10)}&\bf{(58,6)}&\bf{(54,12)}&\bf{(52,14)}&\bf{(52,12)}&\bf{(54,6)}&\bf{(46,14)}&\bf{(40,18)}&(36,18)*& \\
94 &{\bf (54,4)}&\bf{(56,4)}&\bf{(60,0)}&\bf{(56,6)}&\bf{(54,8)}&\bf{(54,6)}&\bf{(56,0)}&\bf{(48,8)}&\bf{(42,12)}&(38,12)*&  \\
92 &\bf{(48,8)}&\bf{(50,8)}&\bf{(54,4)}&\bf{(50,10)}&\bf{(48,12)}&\bf{(48,10)}&\bf{(50,4)}&(42,12)* &  &  &  \\
90 & {\bf(44,8)}&\bf{(46,8)}&\bf{(50,4)}&\bf{(46,10)}&\bf{(44,12)}&(44,10)*&(46,4)* &(38,12)*&  &   &     \\
88 &(42,4)* &(44,4)* &        &        &        &        &        &        &        &        &         \\
 
% \hline
\end{tabular}
\end{table*}

%%%%%%%%%%%%%% TABLE 3  %%%%%%%%%%%%%%%%%%%%%%%%%%%%%%%%%%%%%%%%%%%%%%%%%%%%%
\begin{table*}[htb]

\caption{Same as Table II, but for nuclei with both protons and neutrons  in the 50-82 shell.  Based on Ref. \cite{PRC2}.
}

\bigskip

\begin{tabular}{ r | r r r r r r r r r r r r  }

             & Ba    &  Ce    &  Nd    &  Sm    &   Gd   &  Dy    &  Er    &  Yb    &  Hf    &   W    &   Os   &  Pt     \\
  $Z$         &  56   & 58     & 60     & 62     & 64     & 66     & 68     & 70     & 72     & 74     & 76     & 78      \\

\hline

78 &        &        &        &         &          &        &(18,14) &(20,8)  &$\underline{(12,16)}$&$\underline{(6,20)}$&$\underline{(2,20)}$& $\underline{(0,16)}$ \\
76 &        &(20,16)*&(22,16)*&(26,12)* &(22,18)* &(20,20)* &(20,18) &(22,12)&$\underline{(14,20)}$&$\underline{(8,24)}$&$\underline{(4,24)}$&$\underline{(2,20)}$ \\
74 &(24,12)*&(24,16)* &(26,16)* &(30,12)* &(26,18)* &\bf{(24,20)} &(24,18)&(26,12)&$\underline{(18,20)}$&$\underline{(12,24)}$&$\underline{(8,24)}$&$\underline{(6,20)}$ \\
72 &(30,8)* &(30,12)* &\bf{(32,12)}&\bf{(36,8)}&(32,14) &(30,16)&(30,14)&(32,8)&(24,16)&$\underline{(18,20)}$&$\underline{(14,20)}$ &$\underline{(12,16)}$ \\
70 &(38,0)*&\bf{(38,4)}&\bf{(40,4)}&\bf{(44,0)}&(40,6)&(38,8)&(38,6)&(40,0)&(32,8)&(26,12)&(22,12) &(20,8)  \\
68&\bf{(36,6)}&\bf{(36,10)}&\bf{(38,10)}&(42,6)&(38,12)&(36,14)&(36,12)&(38,6)&(30,14)&(24,18)&(20,18)&(18,14) \\
66 &\bf{(36,8)}&\bf{(36,12)}&(38,12) &(32,8)&(38,14)&(36,16)&(36,14)&(38,8)&(30,16)&(24,20)&(20,20)& (18,16)  \\
64 &\bf{(38,6)}&\bf{(38,10)}&(40,10)&(44,6) &(40,12)&(38,14)&(38,12)&(40,6)&(32,14)&(26,18)&(22,18)& (20,14)\\
62 &\bf{(42,0)}&(42,4) &(44,4) &(48,0) &(44,6) &(42,8) &(42,6) &(44,0) &(36,8) & (30,12) &(26,12) & (24,8) \\
60 &(28,4) &(38,8) &(40,8) &(44,4) &(40,10) &(38,12) &(38,10) &(40,4)&(32,12) &(26,16) &(22,16) &(20,12)  \\
58 &(36,4) &(36,8) &(38,8) &(42,4) &(38,10) &(36,12) &(36,10) &(38,4) &(30,12) &(24,16) &(20,16) &(18,12)  \\
56 &(36,0) &(36,4) &(38,4) & (42,0) &(38,6) & (36,8) & (36,6) & (38,0)& (30,8) & (24,12) & (20,12) & (18,8) \\
 
 %\hline
\end{tabular}
\end{table*}

\end{document}